# Scale invariability

Mensur Omerbashich

*Physics Department, Faculty of Science, University of Sarajevo, Zmaja od Bosne 33, Sarajevo, Bosnia*
*Phone +387-63-817-925, Fax +387-33-649-359, E-mail: momerbasic@pmf.unsa.ba; CC: omerbashich@gmail.com*

I recently demonstrated that the Earth is a mechanical oscillator in which springtide-induced magnification of all-masses' resonance forces tectonics. I here generalize this georesonator concept so to make it apply to any body anywhere in all the universes, and at all times. It turns out that there is no distinction between physics at intergalactic, mechanist, quantum, and smaller scales. Instead of being a constant of proportionality of physics at all scales, $G$ is a parameter of most general form: $G = s \, e^{\wedge}2$, nonlinearly varying amongst different scales $s$. The so-called scale variability of physics but not of $G$, imagined as such by Planck and Einstein, is due to springtide-induced extreme resonance of Earth masses critically impeding terrestrial experiments for estimating $G$, while providing artificial settings for quantum experiments to trivially "work". Thus the derived equation is that of levitation. Reality is a system of near-infinitely many magnifying oscillators, where permanent energy decay of all oscillation forbids constancy of known "physical constants". This hyperresonator concept explains the magnetism (as every forced oscillator's feature), and the gravitation (as forward propagation of mechanical vibrations along the aether i.e. throughout the vacuum structure). To test my claim I propose a Space mission to collect on-site measurements of eigenperiods of the Sun, its planets, and their satellites. The levitation equitation enables propulsionless Space travel via gravity sailing: Space vehicle's hull ought to be engineered so as to automatically adjust its grave mode, to the vehicle's instant gravitational surroundings, akin to trout up-swimming.

## Introduction

Resonant magnification of Earth's total-mass's (mostly the mantle) oscillation causes tectonics and related phenomena on a Moon-forced Earth [1]. Specifically, the total-mass Earth is a forced mechanical oscillator [1], first proposed by Tesla [2], in which the damping force is proportional to the velocity of the body [3], the total-mass's vibrational grave mode $_0P_2$' makes the system normal frequency $\omega_o$, and the gravimeter-measured lunar synodic period $T_{syn}$' makes the system forcing frequency $\omega$, so the states of a maximum oscillation magnification and a maximum stored potential energy, as well as the body's magnetic field, occur [4]. The georesonator concept explains why atmospheric data behave as seismic precursors [5].

In order to examine universality of the georesonator concept, let us assume it valid everywhere and at all times. Assume also that the Planck's and Einstein's presumed realities in both scaling directions beyond (our) reality (i.e., the reality which holds true on mechanist scales; i.e., *undefined totality*) do not hold any more so there exists also a unifying relationship between their physics too – along the mechanist v. quantum scales. Here, I use $G$ to quantify physics. Then in order to search for variations between the values of $G$ along the two scales, while proposing that the georesonator be the only fully known aspect of the reality, I extend my georesonator concept in both directions beyond mechanist scales, while allowing for $G$ to be constant on a given scale albeit not necessarily universally constant. This concept is at least as general as the concept by Planck and Einstein.

I use the results (the lunar synodic forcer of the total-mass Earth, of 14.7655 days; the normal period of the total-mass Earth, of 3455 s) obtained in Gauss-Vaniček variance-spectral analyses of the Canadian superconducting gravimeter's (SG) 1 Hz decadal observations of mass acceleration [1].

## Generalization

The main problem with the Planck's and Einstein's capital theories (quantum mechanics, special and general relativity, respectively) is that they assume some imagined realities beyond obvious reality – at much smaller and at much larger scales respectively, while assuming that $G$ remains constant no matter the reality. In such an arbitrary and extremely speculative setup, any experiments to prove (or not) the setup clearly stand little chance for succeeding. Consequently, one century later, neither has any conclusive proof of the two theories been demonstrated nor a connecting relationship between the two theories found.

Contrary to the Planck-Einstein setup in which ideas ("thought experiments") come before or even without data, I focus on the Earth data only, while letting $G$ unconstrained. I thus generalize the georesonator concept by (i) making no assumptions on $G$, and (ii) tacitly assuming that all gravitational considerations performed in the vicinity of the Earth, such as those aimed at determining $G$ [6], had failed to account for resonant magnification of total-mass's (gravity) oscillations during those considerations.

Then in the limiting case of such a georesonator's maximum magnification given by $M(\omega) = 2Q^2 / \sqrt{4Q^2 - 1}$ (where the spectral spread of the system response about the resonant frequency $\omega = \omega_{max}$ equals $Q = 1/\sqrt{2 - 2(\omega/\omega_o)^2}$) [4], an over-scaling (numerical) relationship is asserted between physics at the Newtonian v. Planck scales, i.e., between the Newtonian constant of gravitation (or: constant of proportionality of physics at mechanist scales), $|G_{Newton}| = (6.6742 \pm 0.001) \cdot 10^{-11}$, c.f. NIST.GOV (2003 estimate), and Newtonian constant of gravitation over $\hbar c$ (or: constant of proportionality of physics at quantum scales), $|G_{Planck}| = (6.7087 \pm 0.001) \cdot 10^{-39}$, c.f. NIST.GOV (2003 estimate), as:

$$G_{Newton} \pm \varepsilon_1 \to G_{Planck} \cdot c^3 \cdot \frac{\omega_o}{\omega_{max}}\bigg|_{Earth-Moon}, \quad \varepsilon_1 \in \Re \quad (1)$$

suggesting there is no real distinction between physics at the mechanist v. quantum scales. If the relationship (1) is real, its analytical uncertainty $\varepsilon_1$ depends only on the uncertainties of $G$ and $\omega_o$. Let us examine this.

As there is no dependence on time $t$ in (1), there is no need for invoking the Riemannian geometry either. The existence of frequency-space approaches to finding a unifying theory should be realistic if non-relativistic quantum mechanics alone is used – say by equating the concepts of Earth mass and charge where the charge could be subject to lunar forcing, just like all the masses comprising the Earth charge are. For instance, Tesla [2] has maintained that reality is describable within the frequency space alone, and, consequently, permanent vibrations (revolutions) of all matter, regardless of the scale, were in fact linearly misconstrued by his predecessors to mean permanent "free falling" of all matter. For instance, in order to abridge this conceptual gap (i.e., in order to complete his theory), Newton had attached units to his constant of proportionality, $G$. As viewed in the herein proposed *hyperresonator (mass all-resonators) concept*, such a mathematical lock occurred in order for the Newton's (Einstein's) linear misconception of our frequency-space and non-obvious reality to be satisfied.

So regarded as a ratio, $G$ is a quantity that carries no units. In the same sense, $c$ represents a unitless ratio as well (the units in the latter are derived anyway). Pragmatically, this is the only way for physical units to "agree" with the above stated assumptions on the georesonator being a concept even more general than the Planck's and Einstein's presumably general theories were.

Then given the SG-measured value $\omega_o$', the following numerical relationships *naturally hold* (i.e., physics of mass-resonators reality at an appropriate scale tends to be a natural function) everywhere in the vicinity of the Earth; as in (1), units are obviously of no concern here:

$$G_{Newton} \cdot c \cdot \frac{\omega_o}{\omega_{max}}\bigg|_{Earth-Moon} \to e^2 \quad (2)$$

with ±0.14% analytical uncertainty, and

$$G_{Planck} \cdot c \cdot \frac{\omega_o}{\omega_{max}}\bigg|_{Earth-Moon} \cdot 10^{28} \to e^2 \quad (3)$$

with ±0.40% analytical uncertainty. From expressions (1) - (3) follows the refinement of the resonance (ratio) of the Moon-forced Earth's total mass, as:

$$\frac{\omega_o}{\omega_{max}}\bigg|_{Earth-Moon} \cdot 10^{-28} \to c^{-3} \quad (4)$$

with ±0.07% analytical uncertainty.

The analytical uncertainties in expressions (2) - (4) are presumably due to the uncertainty of $G$ and $\omega_o$. To explore this, let us insert $T_{max}$' = 14.7655 days into (4), obtaining for the certainty range:

$$T_o\big|_{Earth} \in [3419\,s,\ 3455\,s]. \quad (5)$$

Here the intensity of the interval (5), of $|T_o| = 18$ s, reflects the higher-order terms in the theoretical value [7]: $T_{syn}$ [m.s.d.] = $29.5305888531 + 2.1621 \cdot 10^{-7} \cdot T_{t.d.t} - 3.64\,10^{-10} \cdot T^2_{t.d.t}$, where $T_{t.d.t}$ [JC] = (JD−2451545)/36525 [7]. The same intensity also reflects the 17 s accuracy of lunar synodic period's observations and hence of the 8 s filtering step used [1] too. Importantly, the above intensity falls within certainty limits of SG-observed $\omega_o$' [1].

Of all the possible values from the above interval (5), only the upper-limit value is of interest here, as it is precisely this value that corresponds to the oscillation of the Earth's total mass [1]. So:

$$T_o\big|_{Earth} = 3455\,s. \quad (6)$$

Inserting (6) back into the relationship (1) gives $2 \cdot 10^{-4}$ for the uncertainty of (1), obviously reflecting only the higher-order terms in the theoretical value $T_{syn}$ [m.s.d.] [7], and the NIST relative standard uncertainty for $G_{Newton}$, of $1.5 \cdot 10^{-4}$. Thus the relationship (1) is verified within reason.

It is now justified to write the full relationship, using (6) for the mechanist scale, as:

$$\frac{\omega_o}{\omega_{max}} = C(\tau; T_{t.d.t.}), \quad C \in \Re, \quad C\big|_{Earth-Moon} = 369.2443415 \quad (7)$$

$$\left. \begin{array}{l} G_{Newton} \cdot c \cdot C^{I} \to e^2 \pm 0.006\% \\ \\ G_{Planck} \cdot 10^{28} \cdot c \cdot C^{II} \to e^2 \\ \\ \dfrac{G_{Newton}}{G_{Planck}} \cdot \dfrac{1}{C^{III}} \to c^3, \end{array} \right\} \quad (8)$$

where the Roman superscript in the shorthand $C$ indicates the type of the oscillator of interest, with its own $\tau$-temporally varying resonance. Specifically:

$$\left. \begin{array}{l} C^{I} = \dfrac{\omega_o}{\omega_{max}}\bigg|_{Earth-Moon} \\ \\ C^{II} = C^{I} \pm \varepsilon_2; \quad \varepsilon_2 \in \Re \wedge \varepsilon_2 \ll \varepsilon_1 \\ \\ C^{III} = C^{II} \pm \varepsilon_3; \quad \varepsilon_3 \in \Re \wedge \varepsilon_3 \equiv f(\varepsilon_1, \varepsilon_2; \tau) = \\ \\ = \dfrac{\varepsilon_1}{\varepsilon_2} + \ldots \wedge \varepsilon_3 \gg \varepsilon_1. \end{array} \right\} \quad (9)$$

Note the excellent agreement for the mechanist scales – first of the expressions (8). In the above, the notion of scales is induced such that the uncertainty $\varepsilon$ with which the proposed concept fits the natural relationship on a given type of scales – represents that type of scales.

If the reality is a hyperresonator of masterfully (i.e., with an extremely slow decay) tuned X number of mass resonators (from say a string and under, to the Universe and beyond), where X is an excruciatingly excruciatingly (…) large number but not infinity, then the above-obtained relation amongst the uncertainties

$$\varepsilon_3 \gg \varepsilon_1 \gg \varepsilon_2 \quad (10)$$

indicates that there is no real distinction (meaning: all differences are fully describable within the real, mechanist domain alone) between physics at mechanist (represented by $\varepsilon_1$) v. quantum scales (represented by $\varepsilon_2$). To examine the quality of the estimate of (6), let us substitute equality $C^{III} = C^{II} = C^{I}$ into the remaining two of the expressions (8):

$$\left. \begin{array}{l} G_{Newton} \cdot c \cdot C^{I} \to e^2 \pm 0.006\% \\ \\ G_{Planck} \cdot 10^{28} \cdot c \cdot C^{I} \to e^2 \pm 0.252\% \\ \\ \dfrac{G_{Newton}}{G_{Planck}} \cdot \dfrac{1}{C^{I}} \to c^3 \pm 0.001\%, \end{array} \right\} \quad (11)$$

This made the analytical uncertainty in case of quantum scales somewhat improve compared to (8) albeit not drastically. Thus according to the proposed concept, an inequality $C^{II} \neq C^{I}$ holds, meaning the first two types of scales bear different but similar properties. However, in case of so-called "relativistic" (intergalactic-to-interuniverse) scales (represented by $\varepsilon_3$), all of the analytical uncertainty practically vanished compared to (4), and equality $C^{III} = C^{I}$ now holds. Hence, according to the proposed concept, physics at the "relativistic" scales does not differ from physics at the mechanist scales.

Then the intergalactic scales can be entirely described within the realms of the two other types of scales too, while according to (8) scales smaller than quantum scales are not forbidden. This suggests that the true nature of reality lays in the harmony of all oscillation as proposed by string theory, and not in simplistic sums of infinitely many cases of "free fall", regardless of the curvature of Riemannian (space-time) representations of reality. (An analogue would be a straight line being just an infinitesimally small segment of a circle.) To arrive at this result a common principle at all scales — the resonant property (7) of gravitation — was used.

Perceptions contrary to the proposed concept, such as the Einstein's general relativity with its claimed universality, appear to stem not from real distinctions but from remarkably different values of uncertainties in physical considerations carried out at the three classical types of scales.

Finally, the natural relationship for physics at the mechanist and quantum scales reads, cf. (11):

$$\left. \begin{array}{l} G_{Newton} \cdot c \cdot C^{I} = e^2 \\ \\ G_{Planck} \cdot 10^{28} \cdot c \cdot C^{II} = e^2 \pm \varepsilon_2 \\ \\ \dfrac{G_{Newton}}{G_{Planck}} \cdot \dfrac{c^{-3}}{C^{I}} = 1. \end{array} \right\} \quad (12)$$

**Discussion**

A simple concept is proposed based on variance spectra of decadal oscillation variations of total-mass Earth, as recorded by a superconducting gravimeter. If the real Earth could be shown to satisfy the concept to a reasonably good approximation, then the earthquake forecasting would become a straightforward task [1]. This fitting of the real Earth to the proposed concept could be attempted by using trial-and-error approach, or by computing the World Geoid where $C$ in (7) on mechanist scales is a function, i.e., the first of Eqs. (12). In order to show that the above concept holds everywhere on Earth and at all times, the proposed concept must be applicable to a realm that is beyond that of the Earth herself. In order to ultimately demonstrate that reality is a set of entangled systems of systems of systems of (...) of mechanical oscillators, all one needs to do is to find the unifying relationship between physics at all scales.

With that in mind, hyperresonator concept was tested using the only form in which such a hypothetical concept could reveal itself to us on Earth as absolutely measurable − that of the georesonator. This concept was substantiated by the superconducting gravimeter as the humankind's most accurate instrument. Note that this concept requires permanent decay of all energy (permanent oscillation dampening), where no constants are possible. Nor is the Einstein's irrational equality between time and clocks (i.e., atomic orbital periods) allowed either. In the same sense, so-called "general relativity" merely represents an alternative (and cumbersome) way of describing the dynamics of $\varepsilon_1$ and $\varepsilon_2$ reality as bounded by three only dimensions. Note that (10) says nothing of the scales which are smaller than $\varepsilon_2$, or of to them alleged higher dimensions, e.g. by string theory.

Imperfect relativistic explanations such as those on the perihelion advance of the Mercury, or the neutron chain reaction, must not serve the purpose of proving unfounded concepts such as the Einstein's general relativity. On the contrary, observational facts must be used to construct viable theories [8]. According to the herein proposed concept, properly modeling any gravitational-orbital phenomena (including perihelion advance, nutation, etc.), as observed for any object in a Solar system, requires first the knowledge (measurement) of the grave mode of oscillation of all the total-masses which comprise the gravitational environments of both the object of interest and of with it oscillating (about it orbiting) objects. Only in that way would it be possible to examine if the gravitational–electromagnetic environments can be equated using the mechanical-electrical resonance with resonant magnification of mass oscillation as a proposed universal property of gravitation. In brief, to prove (8) would require a Space mission that would set up absolute and superconducting gravimeters on the surface of the objects of interest − the Sun and about it orbiting bodies in case of our Solar system.

If the hyperresonator could be verified in the above and other ways, and hence applied on the real Earth and our Solar system, the totality of all masses in the Hyperverse (and not just its radio-observable part known as the Universe) could be guessed, where resonant magnification of gravitation would have to be properly accounted for in all permutations of all the existing gravitational (tidally locked) systems, regardless of the scale. Our Universe itself could well be entangled into Hyperverse of oscillating universes, with the *matter attraction* at the same time being the *matter vibration* which is perpetually excited and dampened. No so-called physical constant is absolutely constant then, but only on the given scale and epoch.

In the here proposed concept the following relationship then generally holds, cf. (12):

$$G = s \cdot e^2 \qquad (13)$$

or, writing the scaling factor, $s$, fully:

$$\left.\begin{array}{l} G_{Newton} = C^{-1} \cdot c^{-1} \cdot e^2 \\ G_{Planck} = C^{-2} \cdot c^{-4} \cdot e^2, \end{array}\right\} \qquad (14)$$

which, for the Earth-Moon system (Eq. (6) and theoretical value $T_{syn}$ [m.s.d.]), is well satisfied to within the NIST standard uncertainty for $G$. The Planck constant $h$ is already determined analogously to the principle of SG [9]. Thus besides $G_{Newton}$, $h$ too reflects the effect of the Earth total-mass resonance, $C$, as given by Eq. (7). Hence the power of two in $C^{-2}$, Eq. (14), corroborates the proposed concept. Note that according to general understanding the theory of quantum gravity allows for constants other than $c$, $G$, and $h$, such as the here discovered $C$.

If $C$ is constant then it is the only constant, in which case its value (7) is approximate. Then not only the Earth but also every spatially unique segment of the reality would maintain this constant also. The ratio of that segment's (say the outer Universe) grave mode of oscillation, and the orbiting period of another segment (say an outer-outer Universe orbiting about the outer Universe), would be a constant approximately equal to $C$ as found for the Earth, Eq. (7). In case of the Earth, it is the Moon which largely masks the tidal influences of the rest of the Universe on the Earth. In case of our Universe, it is to it neighboring (about it orbiting) Universe which largely masks the tidal influences of the rest of Hyperverse onto our Universe... The above-proposed Space mission to verify (13) would also show whether $C$ is the universal constant or not. In any case, (13) represents the first analytical expression for *G*. Note here that correlations have already been found which relate physical and geometrical astronomical quantities, such as the mass and periods of transiting planets [10].

Generalized form of (14) gives

$$G_{strings} = C^{-3} \cdot c^{-7} \cdot e^2 = 6.74377 \cdot 10^{-67} . \quad (15)$$

Finally, the most general form of (14) reads:

$$G_{hyperverse} = c^2 \cdot e^2 = 6.64095 \cdot 10^{17}, \quad (16)$$

where the 0-th power of *C* means harmony of all oscillation. Non-fractional *C* are obviously forbidden in local real domains. Therefore, further generalizations of (14) are not possible for scales larger than that of (16).

By extension, there is nothing in the proposed concept that would forbid the lattice of to-us strings from being a Hyperverse of even smaller-scales' worlds, and so on. The logic of infinity seems inconceivable to us as the infinity could correspond not to our (linear) perception of the frequency domain but to that entire domain itself, instead. For instance, the idea of universe's cyclic life was already proposed [11], and the here presented concept is the broadest possible generalization of that concept, even if not self-evident. In particular, the three linear dimensions bind consciousness. As a result, we are incapable of relating ourselves to the reality in terms of frequency space, which is then forever hidden (non-obvious) to our senses but not to our instruments or analysis methods such as spectral analyses. Thus the gravitation in the here proposed concept is neither a force nor geometry. Instead, it is an environment (not a field any more) of omnipresent vibration amongst all of interlocked particles and bodies. Furthermore, it does not "act at distance" by illusory "gravity waves". Rather, the gravitation is vibration, and the 'minus' sign in the Newton's gravitational law in fact is wrong and as such rightfully ignored for all purposes.

The classical (Newtonian; Einsteinian) view is that the gravitation can only be attractive, as based on our everyday geocentric experience. However, this only appears so. We must rather get rid of the geocentric view altogether. Then looking from all the points in the outer space simultaneously in the direction of the Earth, since the gravitation is vibration, its influence spreads forward (repulsively) throughout the space and in all directions, from all the points (particle/body/galaxy/universe) that have their own mass/energy manifestation, e.g., own spin. For this, no special particles so-called gravitons need to be invoked; instead, the gravitation is a never-ending influx of the mechanical waves disturbing (i.e., vibrating along) the "aether" i.e. the vacuum structure.

Thus, what is usually referred to as the "Earth gravitation" is in fact the resultant of the aether-disturbing, as of yet immeasurably high-frequency-waves arriving to the Earth from the whole of Hyperverse onto the observer at a point. This means that all the points with own energy release parts of that energy, thereby causing disturbances in the aether (vacuum structure) resulting in waveform deformations of aether's steady states. This disturbance we call the gravitation. An evident modulation of this disturbance occurs during the full and new Moon, when the Moon obscures the normal (line-of-sight-) direction of the disturbances as they are incoming from the whole Hyperverse, with the Moon and the Sun being the largest nearby concentrations of particles [12]. This immediate obstruction makes the vibrational nature of the gravitation obvious on Earth only during the full and the new moon, and during the eclipses. Then alignment of the Sun and the Moon on the same side of the Earth does not act on us via adding of their illusive tidal forces of attraction, but by allowing simultaneously for the vibration of the vacuum structure from the Hyperverse to push on the Earth's masses so to make them bulge out and mostly so then in the direction of the alignment.

In case of the Earth, all particles composing the Earth transfer parts of their energy onto its surroundings, due to the subatomic particles acting as forced mechanical oscillators, just like the Earth-Moon system does. The totality of energy or orbital momenta transfer from all subatomic particles onto their domicile body of mass i.e. a local group of particles such as the Earth, determines that body's grave mode of oscillation via Eq. (7). The higher the concentration of particles at a spatial locality means the more energetic (massive) the observed concentration of particles, say the Earth. It also means a smaller push exerted by the rest of Hyperverse onto the observed body, i.e., the stronger the apparent gravitational environment of the body. Hence, the so-called "central gravity fields" in the Newtonian concept are illusive too. Thus the ratio

$$W_{\text{Earth-Moon}} / W_{\text{Earth}} \quad (17)$$

approximately determines at all times the value of the scaling parameter of physics, *G*, at the appropriate scale. This is written in general form as

$$W_{\text{body orbital resonance}} / W_{\text{body own resonance}} . \quad (18)$$

It is submitted in the above that the gravitation propagates outward via mechanical waves in the medium composed of vacuum structure (aether). In the same sense then, the radiating propagation of the gravitation gives rise to what is generally observed as the expansion of the Universe. The lunar tides on Earth are an obvious manifestation of repulsive propagation of the gravitational mechanical waves inside the aether. Namely, high and low tides, appearing respectively on the sides of the Earth directly facing and directly opposing the Moon, are caused by the Moon's obstructing the balance of overwhelming totality of gravitational waves that emanate from the whole Hyperverse. A tide is not "a pull exerted by one body of mass onto another"; this is a magical rather than a scientific explanation, stemming entirely from Newton's obsession with mysticism. Instead, a tide is an instantaneous imbalance in the totality of mechanical waves arriving from the Hyperverse to the body. *Tidal distortion* then is the result of *gravitational shadowing* occurring when the body's satellite obstructs the locally apparent line-of-sight between the body and the gravitational-mechanical waves arriving from beyond the body's satellite (that is: from a cone with the tip at a point on the body, say location of the observer on Earth, opened to deep space behind the satellite; Fig.1). This is immediately useful for propulsionless Space travel, on crafts made such that their structural grave mode auto-adjusts to changes in the gravitational environment as the craft travels. This is akin of the trout upstream and up-waterfall swimming ability. Thus (13) is the **levitation equation**.

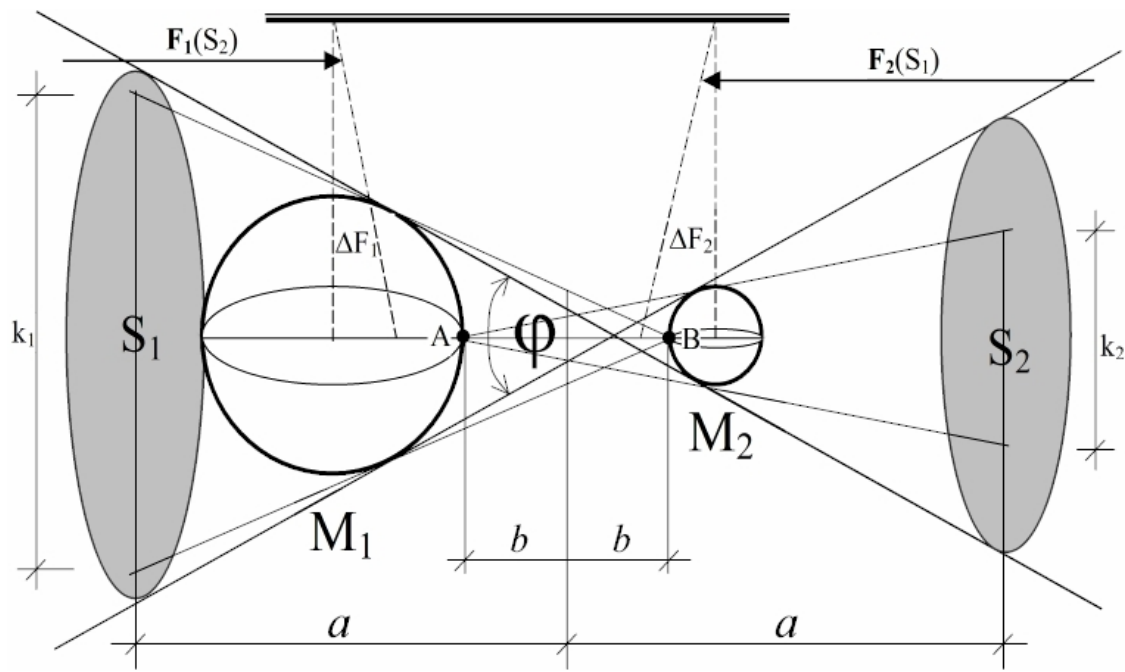

**Figure 1.** *Gravitational shadowing* as felt by the observer at point A (or B) depends on shadow size $k_2$ (or $k_1$). The greater the gravitational shadowing S, the greater the "tidal pull" (in fact: *tidal push*, $\varphi$), **F**. When S-area becomes large enough, a two-body lock occurs. Alternatively, extremely large shadows cause tidal tearing (destruction) of the orbiter $M_2$. The scheme applies to all material bodies (…particles, planets, stars, galaxies…) clustered into systems of mechanical resonators in which kinetic energy gets magnified to the level of an apparent "energy balance" akin of soldiers' step-marching but eventually crashing a bridge.

## Summary

$$\left| G_{Newton} \right|_{NIST} = (6.6742 \pm 0,001) \cdot 10^{-11}$$

$$\left| G_{Planck} \right|_{NIST} = (6.7087 \pm 0,001) \cdot 10^{-39}$$

$w_{0\ Earth}\ /\ w_{Earth\text{-}Moon} = 14.7655\ \text{day}\ /\ 3455\ \text{sec} = C$

$$\left| c^{-1} \right| \cdot C^{-1} \cdot e^2 = 6.6750 \cdot 10^{-11} = \left| G_{Newton} \right|^*$$

$$\left| c^{-4} \right| \cdot C^{-2} \cdot e^2 = 6.7093 \cdot 10^{-39} = \left| G_{Planck} \right|^*$$

Or, written differently:

$$\left| G_{Newton} \right|^* = \left( \frac{\omega_{Earth-Moon}}{\omega_{Earth}} \right)^1 \cdot \frac{e^2}{c} = 6.6750 \cdot 10^{-11} = \left| G_{Newton} \right|_{NIST}$$

$$\left| G_{Planck} \right|^* = \left( \frac{\omega_{Earth-Moon}}{\omega_{Earth}} \right)^2 \cdot \frac{e^2}{c^4} = 6.7093 \cdot 10^{-39} = \left| G_{Planck} \right|_{NIST}$$

Note that the 2006 update of NIST constants improves the estimate of $T_{0\ Earth}$, from a coarse 3455 s, to $3454.8 \pm 0.1$ s.

$$G = s \cdot e^2$$

**Conclusions**

By demonstrating their relationship, I verified the concepts: of mechanist tectonogenesis [1] and of scale-variant *G* and therefore scale-invariant physics. Since the two concepts are theoretically related, computationally correct, and based on some ten billion decade-long gravity observations made by the most precise geophysical instrument, pure coincidences can be safely ruled out. That allows, as a minimum, for unspecified speculations in the general direction to be made. This is the most favorable outcome one could hope for in using seemingly and entirely useless data (geophysical signal *and* noise), and it represents an indication of the correctness of each of the two concepts separately.

Gravity is found herein to be the repulsive mechanical vibration of the aether (i.e., the vacuum's structure, or "fabric"), meaning that aether-detection experiments have yet to become sensitive enough. The benefit of the here derived expression for *G*, which represents the *levitation equation*, is propulsionless Space travel via *gravity sailing*. This can be achieved by engineering Space vehicles such that they automatically adjust their hull's grave mode period so to match the vehicle's immediate gravitational environment, much like when the trout swim upstream and up waterfalls.